\documentclass[10pt, conference]{IEEEtran}

\usepackage{chapterbib}
\usepackage[english]{babel}
\usepackage{graphicx,url}       
\usepackage{mathptmx}           
\usepackage{epsfig}
\usepackage{subfigure}
\usepackage[dvipsnames,svgnames,table]{xcolor}
\usepackage{hhline}
\usepackage{etoolbox}
\usepackage{cellspace}
\usepackage{subfigure}

\usepackage{pdfpages}
\usepackage{url}
\usepackage{color}
\usepackage{xspace}

\usepackage{listings}
\usepackage{tabularx}
\usepackage{array}
\usepackage{pbox}

\usepackage{multirow} 
\usepackage{amsmath}
\usepackage{amssymb}
\usepackage{amsfonts}
\usepackage{amsthm}

\usepackage[latin1]{inputenc}   
\usepackage[T1]{fontenc}
\usepackage{algpseudocode,algorithm,algorithmicx}
\definecolor{MyBlue}{HTML}{5f8dd3}

\definecolor{lightgray}{gray}{0.9}
\definecolor{darkgray}{gray}{0.5}

\usepackage[nolist,nohyperlinks]{acronym}
\usepackage[textwidth=30mm]{todonotes}
\usepackage{soul}
\usepackage{setspace}
\setstcolor{red}
\sethlcolor{SkyBlue}

\theoremstyle{plain}

\newtheorem*{corollary*}{Corollary}

\theoremstyle{definition}

\newtheorem*{definition*}{Definition}

\theoremstyle{remark}

\newtheorem*{remark*}{Remark}

 \begin{acronym}
  \acro{H-CRAN} {Heterogeneous Cloud Radio Access Network}
  \acro{HetNet} {Heterogeneous Network}
  \acro{C-RAN} {Cloud Radio Access Network}
  \acro{RRH} {Remote Radio Head}
  \acro{BBU} {Base-Band Unit}
  \acro{vBBU} {virtual Base-Band Unit}
  \acro{FEC} {Forward Error Correction}
  \acro{HARQ} {Hybrid Automatic Repeat reQuest}
  \acro{LTE-A} {Long Term Evolution Advanced}
  \acro{LTE} {Long Term Evolution}
  \acro{RAN} {Radio Access Network}
  \acro{BER} {Bit Error Rate}
  \acro{Eb/No} {Energy per bit to noise power spectral density ratio}
  \acro{ms} {millisecond}
  \acro{BS} {Base Station}
  \acro{UE} {User Equipment}
  \acro{SNR} {Signal-to-Noise ratio}
  \acro{RAT} {Radio Access Technologies}
  \acro{QoS} {Quality of Service}
  \acro{QoE} {Quality of Experience}
  \acro{GPP} {General Purpose Processor}
  \acro{D2D} {Device-to-Device communication}
  \acro{WiFi} {Wireless Fidelity}
  \acro{F-RAN}{Fog Radio Access Networks}
  \acro{MDC}{Micro Data Center}
  \acro{DC}{Data Center}
  \acro{eICIC}{enhanced Inter-Cell Interference Control}
  \acro{SDN}{Software-Defined Networking}
  \acro{SDWN}{Software-Defined Wireless Networking}
  \acro{km}{Kilometer}
  \acro{CP}{Central Processor}
  \acro{HSPA}{High Speed Packet Access}
  \acro{JT}{Joint Transmission}
  \acro{AWGN}{Additive White Gaussian Noise}
  \acro{Hz}{hertz}
  \acro{QAM}{Quadrature Amplitude Modulation}
  \acro{MIMO}{Multiple-Input Multiple-Output}
  \acro{F-RAN}{Fog Radio Access Network}
  \acro{AI}{Artificial Intelligence}
  \acro{DRL}{Deep Reinforcement Learning}
  \acro{MAS}{Multi-agent System}
  \acro{MCS}{Modulation and Coding Scheme}
  \acro{KB}{Knowledge Base}
  \acro{ASP}{Application Service Provider}
  \acro{CQI}{Channel Quality Indicator}
  \acro{CPU}{Central Processing Unit}
  \acro{FIPA}{Foundation for Intelligent Physical Agents}
  \acro{ACL}{Agent Communication Language}
  \acro{VM}{Virtual Machine}
  \acro{BILP}{Binary Integer Linear Programming}
  \acro{mbps}{Megabit per second}
  \acro{bps}{Bits per second}
  \acro{CDR}{Call Detail Record}
  \acro{3GPP}{Third Generation Partnership Project}
\end{acronym}

\usepackage{comment}
\usepackage{todonotes}
\usepackage{amsmath,amssymb,amsfonts}
\IEEEoverridecommandlockouts
\usepackage{xcolor}
\usepackage{color}
\usepackage{caption}

\usepackage{newtxtext,newtxmath}

\newcommand{\linebreakand}{%
  \end{@IEEEauthorhalign}
  \hfill\mbox{}\par
  \mbox{}\hfill\begin{@IEEEauthorhalign}
}

\title{Impact of User Privacy and Mobility \\ on Edge Offloading}

\author{

	\IEEEauthorblockN{Jo\~ao Paulo Esper\IEEEauthorrefmark{1}, Nadjib Achir\IEEEauthorrefmark{2}\IEEEauthorrefmark{3}, Kleber Vieira Cardoso\IEEEauthorrefmark{4}, Jussara M. Almeida\IEEEauthorrefmark{1}}
	\IEEEauthorblockA{\IEEEauthorrefmark{1}Universidade Federal de Minas Gerais, Brazil. \IEEEauthorrefmark{2}Universit\'{e} Sorbonne Paris Nord, France.\\ \IEEEauthorrefmark{3}Inria, France. \IEEEauthorrefmark{4}Universidade Federal de Goi\'{a}s, Brazil.\\
	E-mail: \{joaopauloesper, jussara\}@dcc.ufmg.br\IEEEauthorrefmark{1}, nadjib.achir@inria.fr\IEEEauthorrefmark{3}, kleber@inf.ufg.br\IEEEauthorrefmark{4}}
	
}

\makeatletter
\def\ps@IEEEtitlepagestyle{%
  \def\@oddfoot{\mycopyrightnotice}%
  \def\@oddhead{\hbox{}\@IEEEheaderstyle\leftmark\hfil\thepage}\relax
  \def\@evenhead{\@IEEEheaderstyle\thepage\hfil\leftmark\hbox{}}\relax
  \def\@evenfoot{}%
}
\def\mycopyrightnotice{%
  \begin{minipage}{\textwidth}
  \centering \scriptsize
  Copyright~\copyright~2023 IEEE. Personal use of this material is permitted. Permission from IEEE must be obtained for all other uses, in any current or future media, including\\reprinting/republishing this material for advertising or promotional purposes, creating new collective works, for resale or redistribution to servers or lists, or reuse of any copyrighted component of this work in other works by sending a request to pubs-permissions@ieee.org.
  \end{minipage}
}
\makeatother

\begin{document}
    \maketitle
    
    \begin{abstract}
    
    Offloading high-demanding applications to the edge provides better quality of experience (QoE) for users with limited hardware devices. However, to maintain a competitive QoE, infrastructure, and service providers must adapt to users' different mobility patterns, which can be challenging, especially for location-based services (LBS). Another issue that needs to be tackled is the increasing demand for user privacy protection. With less (accurate) information regarding user location, preferences, and usage patterns, forecasting the performance of offloading mechanisms becomes even more challenging. This work discusses the impacts of users' privacy and mobility when offloading to the edge. Different privacy and mobility scenarios are simulated and discussed to shed light on the trade-offs (e.g., privacy protection at the cost of increased latency) among privacy protection, mobility, and offloading performance.

    \end{abstract}
    
    
    \section{Introduction}

Mobile devices have undergone a significant transformation from small devices with limited capacity to mobile mini-computers, leading to exponential growth in the mobile application markets. However, with this growth comes a sharp increase in application needs in terms of computational resources from applications such as virtual reality (VR), augmented reality (AR), and interactive games \cite{addad2020fast}, which even our modern mobile terminals cannot well fulfill.

To tackle this issue, edge computing was proposed as a technology in which computational resources are moved to the Edge of the network to reduce latency and guarantee the quality of service \cite{abbas2017mobile}. Thus, users may run their applications on the Edge using offloading techniques. This combination provides the user equipment (UEs) with reduced CPU usage, extended battery life, support for more robust and sophisticated applications, and potentially unlimited storage \cite{zhan2020mobility}.

One challenge that offloading algorithms face is related to user mobility \cite{yang2019efficient}. Users who run their applications on the Edge can move freely in the city, which may significantly affect some applications. Thus, adapting to the users' mobility patterns becomes a challenge as Multi-access Edge Computing (MEC) providers want to keep users' Quality of Experience (QoE) while also dealing with the stress of mobility on their infrastructure through complex operations, e.g., task reallocation.

\begin{figure}[httt!]
\begin{center}
 	\includegraphics[width=0.49\textwidth]{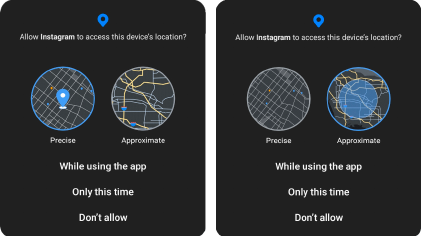}
 	\caption{Precise vs. approximate location on Android 12.}
 	\label{fig:PreciseVsApproximateLocation}
\end{center}
\vspace{-0.75cm}
\end{figure}

Another issue that operators and service providers have recently faced is that users are becoming increasingly aware and concerned about their privacy. As such, governments in various parts of the world responded to this concern with data protection and privacy regulations, e.g., RGPD\footnote{https://eur-lex.europa.eu/eli/reg/2016/679/oj} in Europe and LGPD\footnote{https://www.gov.br/cidadania/pt-br/acesso-a-informacao/lgpd} in Brazil. For example, as illustrated by Figure~\ref{fig:PreciseVsApproximateLocation}, Android 12, released in 2021, changed how user location is accessed, and now users can choose between providing their precise locations or approximate ones.

Privacy concern is an issue for operators and service providers because they need to keep users' privacy and adapt and improve their techniques, e.g., offloading techniques, to maintain competitive performance. In most cases, the more accurate the information you have about the user (e.g., his location), the easier it becomes to predict and anticipate user demand. Yet, most prior studies have evaluated Edge offloading without any consideration of location privacy, especially in complex scenarios composed of multiple applications, each one with distinct requirements, as well as heterogeneous mobility patterns (reflecting, for instance, different mobility types), while analyzing how each of these factors (and combinations of them) may affect the offloading performance.

Our goal is to fill this gap by investigating the impact of user mobility and privacy when offloading to the Edge in multi-application and multi-mobility scenarios. To that end, we employ a combination of state-of-the-art techniques/tools related to mobility and privacy to simulate a large set of scenarios with different user privacy and mobility patterns. As such, our goal is {\it not} to propose a new privacy or security technique but rather to offer insights (e.g., how slow mobility types can suffer more with privacy than fast ones) into the trade-offs between user privacy, mobility, and computational offloading performance.

The rest of the paper is structured as follows. Section~\ref{secII} presents related work. Section~\ref{secIII} describes the system model and problem description. Sections~\ref{secIV} and \ref{secV} discuss our evaluation methodology and main results, respectively, while Section~\ref{secVI} offers conclusions and directions for future work.
    \section{Related work}\label{secII}

Prior work mostly related to our effort tackles (combinations of) aspects related to user privacy, mobility, multi-application scenarios, edge computing, and offloading. However, many of them only considered a few of these aspects. For example, in \cite{he2017privacy}, authors discussed the privacy issues of task offloading with MEC from two perspectives, location privacy, and usage privacy. The authors proposed a Markov decision process to improve latency and energy consumption while keeping pre-specified levels of privacy, but no mobility analysis was offered. Authors in~\cite{sun2020reducing} focused on minimizing offloading latency considering user mobility and movement for AR users but did not consider privacy. In turn, authors in~\cite{wei2021uav} focused on task offloading in a scenario of unmanned aerial vehicles with limited energy and computing capacities. The paper focused on privacy but from a computation offloading preference leakage perspective, not focusing on user location privacy.

In contrast, two recent studies have jointly explored most of those aspects. In \cite{peng2022mobility}, the authors proposed a mobility and privacy-aware offloading meta-heuristic method for a particular scenario, namely, AR applications that deal with patients' private information in healthcare systems. Instead, we here consider scenarios where requests from diverse applications compete to be able to offload their requests. In \cite{pang2022towards}, the authors focused on the privacy leakage issue of MEC offloading, assuming an honest-but-curious server as we do in this paper. They proposed a computational offloading mechanism that provides user privacy while minimizing the total computation cost. However, their work considered homogeneous user mobility patterns, whereas we explored different mobility classes, which, as will be discussed, significantly impact the results.
    \section{System model and problem description}\label{secIII}

Let us consider a MEC environment composed of a set $\mathcal{B}=\{1, 2,..., \textit{B}\}$ of base stations (BSs) and a set $\mathcal{M}=\{1, 2,..., \textit{M}\}$ of MEC Hosts (MHs), as illustrated in Figure~\ref{fig:systemOverview}. Any BS can reach and offload user applications to any MH of choice. Each BS $i$ (MH $j$) has a throughput capacity of $\mathcal{T}^{BS}_{i}$ ($\mathcal{T}^{MH}_{j}$). We consider a set $\mathcal{U}=\{1, 2, ..., \textit{U}\}$ of users, each one modeled by a mobility type, an application, and a privacy level. The user mobility type can be car passenger, bus passenger, or pedestrian. The application can be video streaming, AR, or VR. Finally, the privacy level can be \textit{none}, \textit{medium}, or \textit{high}. Each application has particular requirements in terms of network resources, which are expressed as:

\begin{figure}[htt!]
\begin{center}
\centering
 	\includegraphics[width=0.49\textwidth]{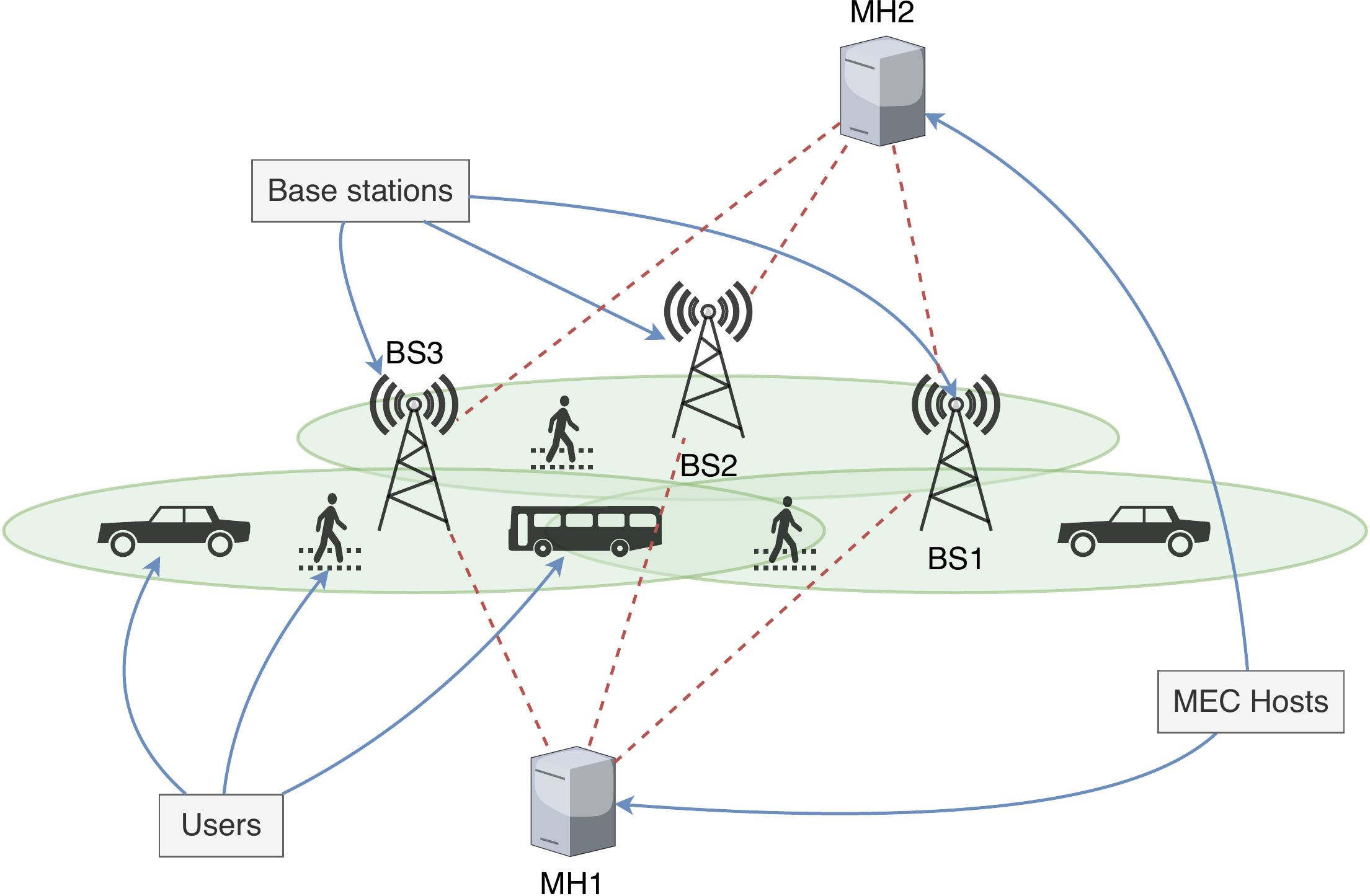}
 	\caption{System overview.}
 	\label{fig:systemOverview}
\end{center}
\vspace{-0.3cm}
\end{figure}

\begin{itemize}
    \item $\mathcal{A}^t$: the throughput demand (i.e., demanded traffic volume), measured in megabits per second (Mbps).
    \item $\mathcal{A}^l$: the latency demand (i.e., maximum accepted latency), measured in milliseconds (ms).
\end{itemize}

The throughput demand is met if there are enough network resources available at the selected MH $j$ (i.e., $\mathcal{T}^{MH}_{j} \geq \mathcal{A}^t$) and if the throughput between the UE and the BS the user is associated to is greater or equal than the throughput demand. The throughput between UE-BS is computed using Shannon's capacity formula~\cite{stallings2017data}. If the total throughput from the UEs connected to BS $i$ exceeds $\mathcal{T}^{BS}_{i}$, BS $i$'s capacity is distributed among them following the proportional fairness method \cite{kim2005proportional}.

The latency demand is met if the latency between the BS the user is associated with and the MH selected to handle his request is less than or equal to the latency demand. The latency between BS-MH in this work is calculated based on the distance between the BS and the MH.

This work assumes that the UE is always associated with the closest (Euclidean distance) BS. Once the UE is associated, it requests the network provider to offload an application. The network provider will then report the UE's position to the MEC provider, which, in turn, will assign the closest MH to the BS the UE is associated with to handle the user's request, i.e., offload the application. Finally, if the requirements of the user's application are met, it is offloaded to the MEC system, consuming $\mathcal{A}^t$ resources from the MH $j$ it was allocated to.

\begin{figure}[httt]
\begin{center}
 	\includegraphics[width=0.49\textwidth]{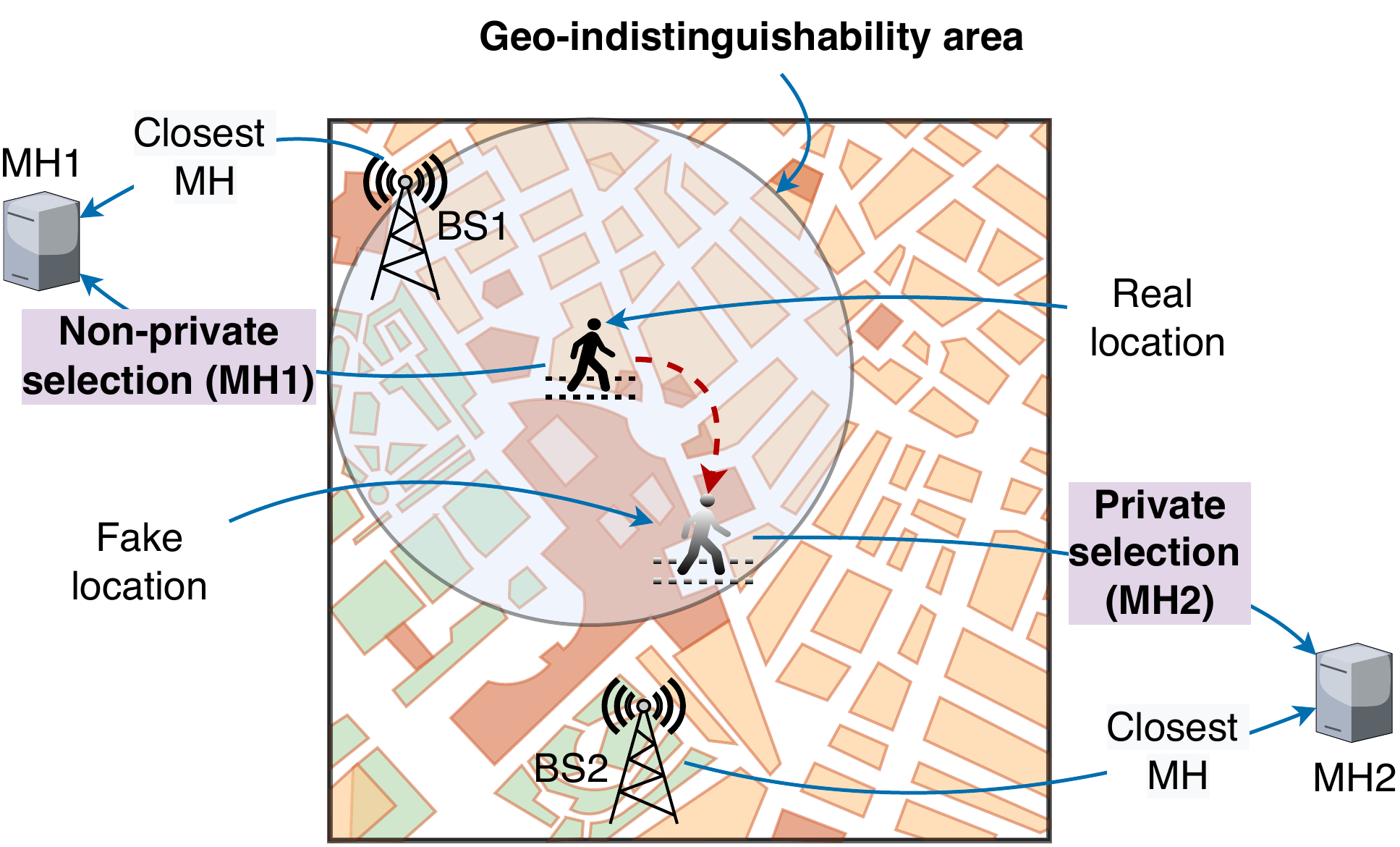}
 	\caption{Privacy overview.}
 	\label{fig:privacyOverview}
\end{center}
\vspace{-0.4cm}
\end{figure}

Figure \ref{fig:privacyOverview} illustrates how we model privacy in our system. We assume the existence of two infrastructure entities: i) the network provider; responsible for the BSs, communication, routing, etc., and ii) the MEC provider; responsible for the MH computational resources. {\it We aim to protect user location from the MEC provider's perspective}, as trying to mask (or fake) the user's real location from the network provider is a different challenge since the UE is associated with a BS.

In Figure \ref{fig:privacyOverview}, we represent two different MH selections, a \textit{non-private} selection and a \textit{private} selection. A \textit{non-private} selection, illustrated by the pedestrian in the center of the circle, occurs when the user is {\it not concerned with his location privacy} (privacy level set to {\it none}). In this case, the network provider will report the user's real location to the MEC provider, and the MEC provider will assign MH1, which corresponds to the closest MH to the BS to which the user is connected. MH1 will then handle the user's request.

A {\it private} selection, in turn, occurs when the user is concerned about his location privacy, which corresponds to a privacy level set to {\it medium} or {\it high}. In this case, the network provider knows the user's real location. Yet, it will apply geo-indistinguishability \cite{andres2013geo}, a state-of-the-art technique based on differential privacy, to generate a fake location, represented by the pedestrian near the circle's edge (see further discussion below). The network provider reports to the MEC provider the user's fake location (not the real one). The MEC provider, in turn, will assume the user is connected to the BS at the bottom of the figure and will assign MH2 to it, as it is the closest MH to that particular BS. MH2 will then handle the user's request. Hence, the user will gain privacy but he may lose performance as his application may no longer be offloaded to the MH that is closest to the BS he is actually connected to. This may affect application latency, resulting in QoE degradation or even failing the application demands, leading to a denied request.

To check if the application latency demand is met, the MEC provider asks the network provider to measure the latency between the selected MH and the BS to which the user is connected. The application is offloaded if the obtained latency is less than or equal to the application demand. We note that the MEC provider could ask the network provider to measure the latency between the selected MH to all BSs to gain more information about the user's real location. Yet, it would be cost-inefficient to do that for every user request in a large-scale environment. Also, to efficiently guarantee user location privacy, we assume an honest-but-curious adversary \cite{pang2022towards, paverd2014modelling}, meaning that the MEC provider will not deviate from the defined protocol, even though it attempts to learn all possible information from legitimately received requests.

\begin{figure}[ttt!]
\begin{center}
 	\includegraphics[width=0.49\textwidth]{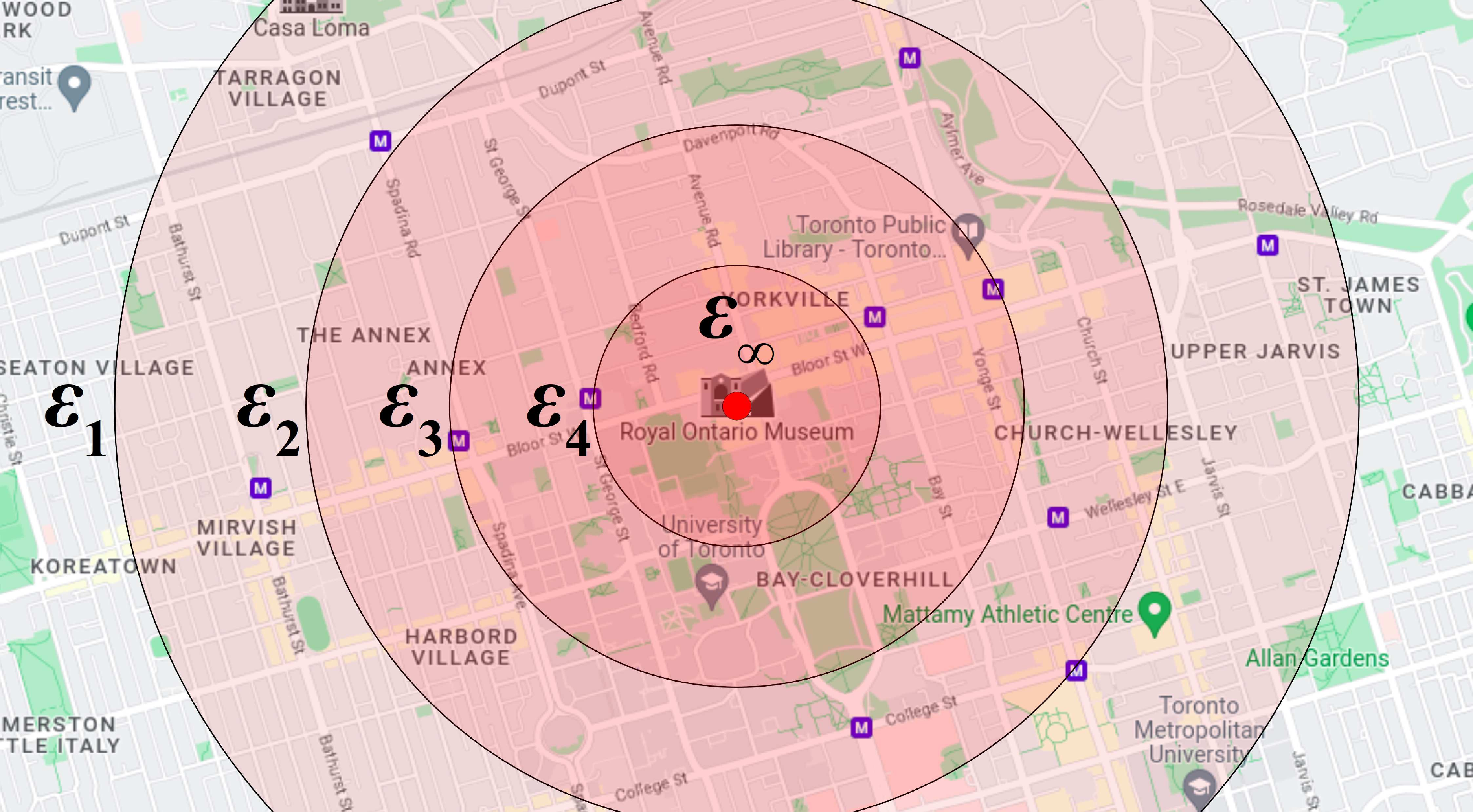}
 	\caption{Geo-indistinguishability mechanism representation, inspired by \cite{andres2013geo}.}
 	\label{fig:geo-indistinguishability}
\end{center}
\vspace{-0.4cm}
\end{figure}

As mentioned, we here use geo-indistinguishability to provide location privacy for the user by producing a new (fake) location. Figure \ref{fig:geo-indistinguishability} illustrates how this method works. Consider a tourist in Toronto, visiting the \textit{Royal Ontario Museum}, who intends to offload his application or do any kind of LBS request. The tourist, concerned with his privacy, does not wish to share his precise location (represented by a red point). Instead, he will share an approximate (fake) location. Geo-indistinguishability works by defining a circle centered around the user's real location and uniformly selecting a new/fake location inside this circle. The circle's radius is computed based on parameter $\varepsilon$. \textbf{Smaller values of $\varepsilon$ result in greater circle's radius}. The new location selection is uniform across the circle. Thus a bigger circle around the user will cover more locations, lowering the probability that real and fake locations are close to each other. Thus, {\it greater privacy levels are achieved by using smaller $\varepsilon$ values} as these lead to greater noise added to fake the user's real position.

Given that the relationship between $\varepsilon$ and privacy is inversely proportional, $\varepsilon = \infty$ is a baseline in this mechanism, and it represents the user's real position, i.e., if $\varepsilon = \infty$, then privacy is non-existent. In Figure \ref{fig:geo-indistinguishability}, five different $\varepsilon$ values are presented, each one next to the circle it generated, such that $\varepsilon_{\infty}$ > $\varepsilon_{4}$ > $\varepsilon_{3}$ > $\varepsilon_{2}$ > $\varepsilon_{1}$. As illustrated, with a larger value of $\varepsilon$, e.g., $\varepsilon_{4}$, geo-indistinguishability has fewer locations to assign a new (fake) position for the tourist, mainly outside of the museum or in nearby streets. In contrast, a smaller value of $\varepsilon$, e.g., $\varepsilon_{3}$, can generate fake positions that $\varepsilon_{4}$ could not, such as the \textit{University of Toronto} or the \textit{Toronto Public Library}.

    \begin{table*}[tttt]
\caption{Simulation parameters.}
\centering
\begin{tabular}{|c|c|c|c|c|c|c|c|c|c|c|}
\hline
\textbf{Runs} & \textbf{Area} & \textbf{Users} & \textbf{Car pas.} & \textbf{Bus pas.} & \textbf{Pedes.} & \textbf{BSs} & \textbf{MHs} & \textbf{BS capacity} & \textbf{MH capacity} & \textbf{Privacy levels} \\ \hline
30 & 4 km$^2$ & 1250 & 400 & 400 & 450 & 475 & 95 & 10 Gbps & 10.41 Gbps & $\varepsilon$ = $\infty$, 0.1, 0.01 \\ \hline
\end{tabular}
\label{tab:simulationParameters}
\end{table*}

\begin{table*}[tttt]
\caption{Applications parameters.}
\centering
\begin{tabular}{|c|c|c|c|c|c|}
\hline
\textbf{Application} & \textbf{Bandwidth req.} & \textbf{Latency req.} & \textbf{\% of cars pas. using} & \textbf{\% of bus pas. using} & \textbf{\% of pedes. using} \\ \hline
Video & 70 Mbps & 10 ms & 70\% & 70\% & 70\%    \\ \hline
AR & 100 Mbps & 30 ms & 15\% & 15\% & 30\%    \\ \hline
VR & 132 Mbps & 14 ms & 15\% & 15\% & -     \\ \hline
\end{tabular}
\label{tab:appsParameters}
\end{table*}

\section{Evaluation methodology}\label{secIV}

This section defines our evaluation methodology by introducing the main parameters used to build our evaluation scenarios. Recall that our goal is to tackle the question: \textit{What are the impacts of user privacy and mobility when offloading to the edge?} Thus, our parameters, summarized in Tables \ref{tab:simulationParameters} and \ref{tab:appsParameters}, relate to mobility, privacy, and (edge) infrastructure and computational resources.

We simulated different user mobility patterns by considering three mobility types: car passengers, bus passengers, and pedestrians. To that end, we used SUMO (\textit{Simulation of Urban MObility})\footnote{https://www.eclipse.org/sumo/}, a widely used mobility simulator. We simulated 30 different runs with SUMO, each with a different seed. Each run lasts for 1 hour, with a temporal resolution of 1 second, and considers a squared area of 4 km$^2$ (2 km * 2 km). We used the population density reported in \cite{zhou2015spatial}, i.e., 312.5 users/km$^2$, resulting in a total of 1,250 users, which were distributed into the three mobility types as follows: 400 car passengers, 400 bus passengers, and 450 pedestrians. We assume that each car has one passenger and each bus has 10 passengers, generating a total of 440 vehicles in the area (400 cars and 40 buses).

To quantify and position BSs and MHs in the simulated area, we used Homogeneous Poisson point processes (HPPP), as done in \cite{zhou2015spatial}. Specifically, we set the HPPP intensity parameter $\lambda$ to 118.75, as reported for `Urban Area 1' in \cite{zhou2015spatial}, leading to 475 BSs, and used $\lambda$=23.75 to generate and position 95 MHs in the simulated area. We assumed that each BS $i$ has a capacity $\mathcal{T}^{BS}_{i}$ of 10 Gbps, as reported by 3GPP~\cite{3gpp2016study}, and each MH $j$ has a capacity of $\mathcal{T}^{MH}_{j}$ of 10.41 Gbps~\cite{spinelli2020toward}.

We defined three user privacy levels by varying the $\varepsilon$ parameter used by the geo-indistinguishability method. As discussed, the smaller the value of $\varepsilon$, the more private the user's (real) position becomes. The first level, referred to as \textit{none}, corresponds to $\varepsilon = \infty$, the case when the user chooses not to have location privacy (user's real and fake positions are the same). It is used here as a baseline for comparison. The other two levels, i.e., \textit{medium} and \textit{high}, are defined by setting $\varepsilon$ = 0.1 and $\varepsilon$ = 0.01 respectively, following a prior study on location privacy in the edge~\cite{qiao2019effective}. 

To make our evaluation more realistic, we also considered the throughput and latency requirements of three different types of applications: video streaming, augmented reality (AR), and virtual reality (VR), as presented in Table \ref{tab:appsParameters}. The video streaming and AR requirements were obtained from \cite{spinelli2020toward}, whereas those for VR were obtained from \cite{lai2017furion}. Following Cisco\footnote{https://www.cisco.com/c/en/us/solutions/collateral/executive-perspectives/annual-internet-report/white-paper-c11-741490.html} and Ericsson\footnote{https://www.ericsson.com/en/mobility-report/reports/november-2019/mobile-traffic-by-application-category} forecasts for network traffic in the following years, we defined that 70\% of the users are consuming a video streaming application. We distributed the rest of the traffic (30\%) evenly between AR and VR. We assume that the VR application does not fit the pedestrian mobility type since they cannot be fully aware of their surroundings if they use a VR headset while walking. Thus, we set that the pedestrians can only use a video streaming or AR application, while other mobility types can use any of the three applications.

We conducted several experiments covering three user mobility patterns, three application types, and three privacy protection scenarios. In short, given an input privacy level ($\varepsilon = \infty, 0.1$ or $0.01$), we ran 30 simulations (by varying the seed) with user population (mobility patterns and applications) as described above, producing a total of 90 different experiments. As discussed next, our results present very narrow 95\% confidence intervals, suggesting high accuracy.
    \section{Evaluation results}\label{secV}

We now discuss our results of the impact of privacy, mobility and application types on the offloading performance.

\subsection{Impact of privacy level}

We start by evaluating the impact of the privacy level, captured by the value of $\varepsilon$, on the acceptance of requests. We do so by comparing the outcome of each user request -- offload was successful or failed -- for each privacy level ($\varepsilon$ equal to $\infty$, 0.1, and 0.01). Recall that an offload request is successful whenever there is enough resource at the selected MH and the user application's bandwidth and latency requirements are met. Considering the totality of user requests issued during a simulation run, we classify them into one of the three scenarios: \textit{Always offloaded}, \textit{Privacy dependent}, and \textit{Never offloaded}. The former corresponds to requests that were successful in all runs, whereas the latter relates to requests that failed in all runs, {\it regardless of the value of $\varepsilon$}. Both categories relate to requests that were {\it not impacted by the particular privacy level adopted}. Requests whose outcome varied depending on the value of $\varepsilon$ used fall into the {\it privacy dependent} category.

\begin{table}[ttt!]
\caption{Impact of privacy on acceptance of requests: fraction of requests in each category and 95\% confidence intervals.}
\centering
\begin{tabular}{ccc}
\hline
\textbf{Always offloaded}    &\textbf{Privacy dependent} &\textbf{Never offloaded} \\ \hline
$56.37\%\pm0.06$               &$30.03\%\pm0.07$             &$13.58\%\pm0.10$            \\ \hline
\end{tabular}
 \label{tab:percentageOfRequestsOffloaded}
\end{table}

Table \ref{tab:percentageOfRequestsOffloaded} presents the average fraction of user requests falling in each category, along with 95\% confidence intervals. As shown, most requests (around 56\%, on average), were offloaded in all privacy levels. This means that, in all these cases, regardless of the privacy level chosen ($\infty$, $0.1$, or $0.01$), the user was always able to successfully offload his desired application. We delve deeper into this result by breaking down these requests by application type. We note that over $98\%$ of the requests for AR (the least bandwidth and latency demanding application) were offloaded in all privacy levels. In contrast, only around $47\%$ of the video requests (the most demanding in terms of latency) and $31\%$ of the VR requests (the most demanding in terms of bandwidth) were always offloaded regardless of the privacy level. Thus, as one might expect, privacy has less impact on offloading requests that impose lower requirements. Yet, we note that even though both application requirements and privacy needs could be successfully fulfilled for those requests, users still paid a price for privacy as the latency of these requests did increase for higher levels of protection (as discussed below).

\begin{figure}[ttt!]
\begin{center}
 	\includegraphics[width=0.455\textwidth]{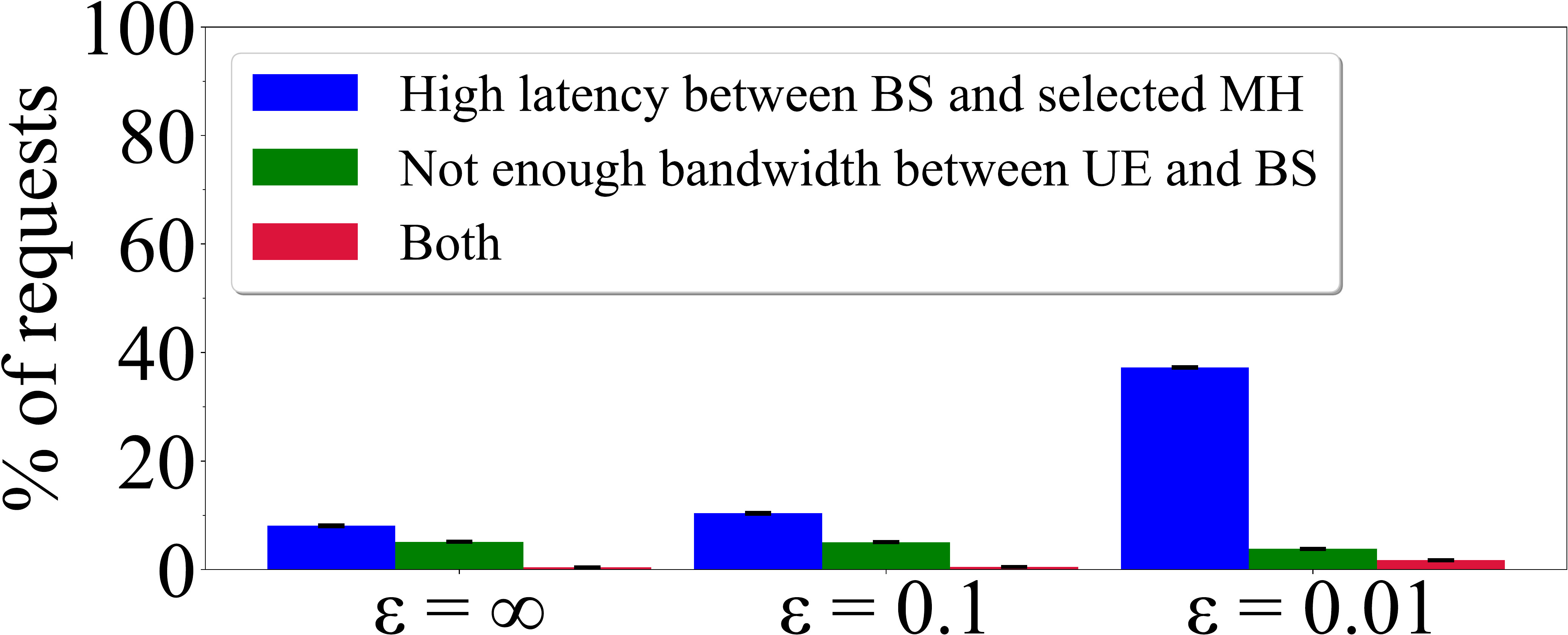}
 	\caption{Reasons for request denials for various privacy levels.}
 	\label{fig:reasonsWhyRequestsWereNotOffloaded}
\end{center}
\vspace{-0.3cm}
\end{figure}

In contrast, around 30\% of the requests fell into the \textit{Privacy dependent} category, implying that the success of the offload depended on the user's privacy level. The vast majority of those requests, or 28\% of all requests, were accepted for privacy levels equal to {\it none} and {\it medium} failing only for privacy level equal to {\it high}. Thus, in total, a user was able to successfully offload his application while still maintaining a {\it medium} privacy protection in roughly 84\% of the time (56\% + 28\% of all requests). A {\it high} privacy level, in turn, is more challenging as it more often leads to request being denied.

We note that around 14\% of the requests were never offloaded, regardless of the privacy level (even for no privacy protection). We then zoom deeper into the reasons why users were not able to offload their applications in each of the privacy scenarios. Recall that the denial of an offload request may happen because\footnote{We note that, for simulation purposes and given the limited area considered, all requests are always associated with a BS. Thus, request denials cannot occur due to the lack of coverage by nearby BS.}: 1) the latency between the BS the UE is associated with and the selected MH is above the application's latency requirement (blue bars); 2) the throughput between the UE and the BS is not enough to meet the application's throughput requirement (green bars); 3) both latency and throughput requirements are not met (red bars).

As shown in Figure~\ref{fig:reasonsWhyRequestsWereNotOffloaded}, for all privacy levels, the first reason -- high latency between BS and MH -- dominates all cases, often occurring for requests to offload video streaming, the most latency demanding application. Moreover, the fraction of requests denied due to this reason increases with privacy protection. This happens because smaller $\varepsilon$ values imply in greater chance that the MEC operator selects an MH that is farther from the BS the UE is associated with (true and fake user locations far from each other), resulting in higher latency.

The other two reasons, in turn, occurred with much lower frequency, often for requests to offload the VR application (with the highest bandwidth demands). We note that the privacy level does {\it not} impact the fraction of requests denied due to lack of throughput between the UE and the BS (reason 2) since there is no privacy from the network operator's standpoint (i.e., user's BS is the same for all privacy levels). Yet, we do observe a small increase in the fraction of requests denied due to reason 3, i.e., both application requirements are not fulfilled, in the most private scenario ($\varepsilon = 0.01$). Such increase is mostly due to the longer latency that result from the higher privacy protection, as discussed above.

\begin{figure}[tt!]
\begin{center}
 	\includegraphics[width=0.455\textwidth]{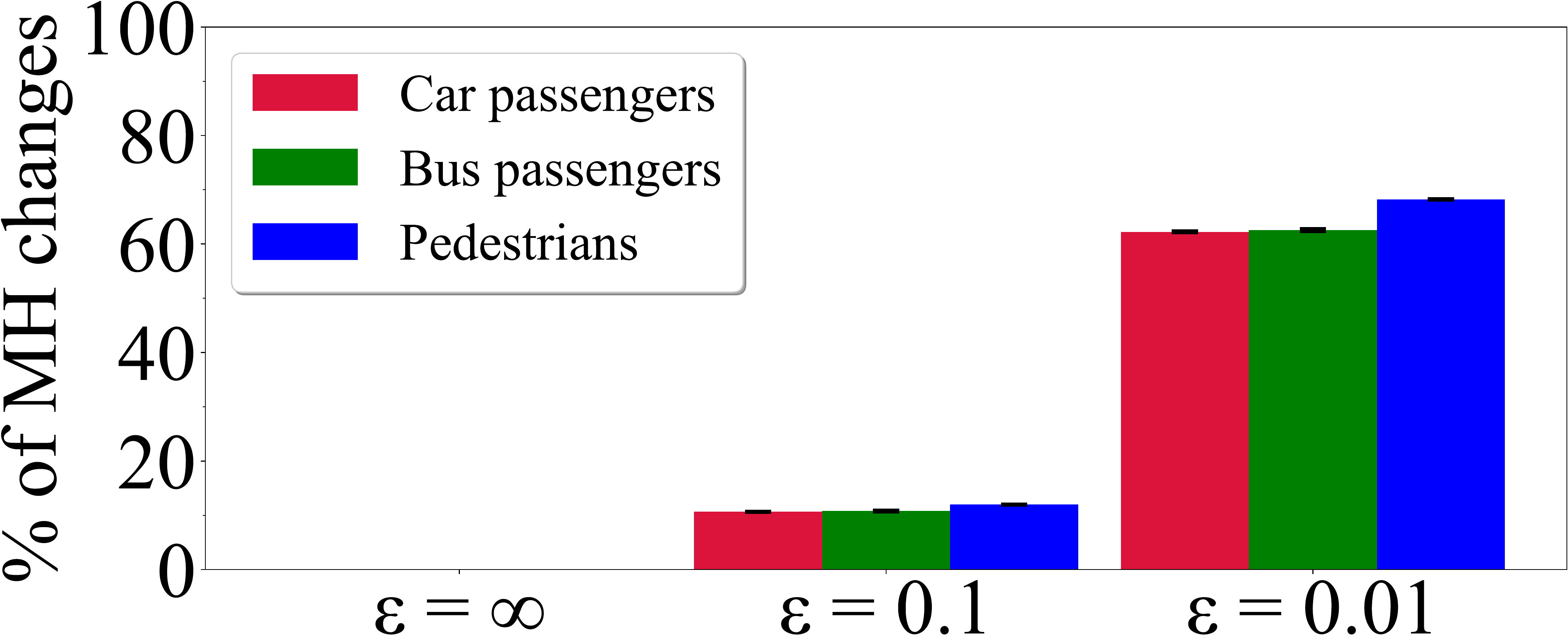}
 	\caption{Impact of privacy and mobility on the MH selection.}
 	\label{fig:mhChanges}
\end{center}
\vspace{-0.3cm}
\end{figure}

\subsection{Impact of mobility type}

We now turn our attention to how privacy impact the MH selection for different mobility types. As 
detailed in Section~\ref{secIII}, the MEC operator selects the MH closer to the BS he thinks the UE is associated with, which might be right or not. Figure~\ref{fig:mhChanges} presents the fractions of MH selections that differ from the ideal one (closest to the true BS) for each mobility type.

Naturally, with no privacy protection ($\varepsilon = \infty$), the selected MH is always ideal. Yet, the fraction of selections resulting in different MHs increases with the privacy level for all three mobility types, reaching at least 60\% for {\it high} privacy. Also, Figure~\ref{fig:mhChanges} shows that the slowest mobility type (pedestrian) is more impacted by privacy than the other two types, reaching 70\% of non-ideal MH selections for the {\it high} privacy level.

Prior work has analyzed the impact of mobile user's speed on offloading performance under privacy protection \cite{pang2022towards}, focusing on a single mobility type. Our result suggests that, in addition to speed, different user mobility patterns, notably \textit{where} and \textit{how} the user moves, may also impact offloading with privacy protection. Pedestrians move only on the sidewalks and crosswalks, while cars and buses move only on the streets. Also, car/bus flow is influenced by traffic jams and traffic lights, while pedestrians move more freely on the sidewalks, hardly ever facing traffic lights on crosswalks. Depending on BSs and MHs limitations, MEC selection may be impacted differently as spatial constraints (sidewalks, crosswalks or streets) of each mobility type can offer an advantage (or disadvantage) once privacy protection is applied.

\begin{figure}[ttt!]
\begin{center}
 	\includegraphics[width=0.455\textwidth]{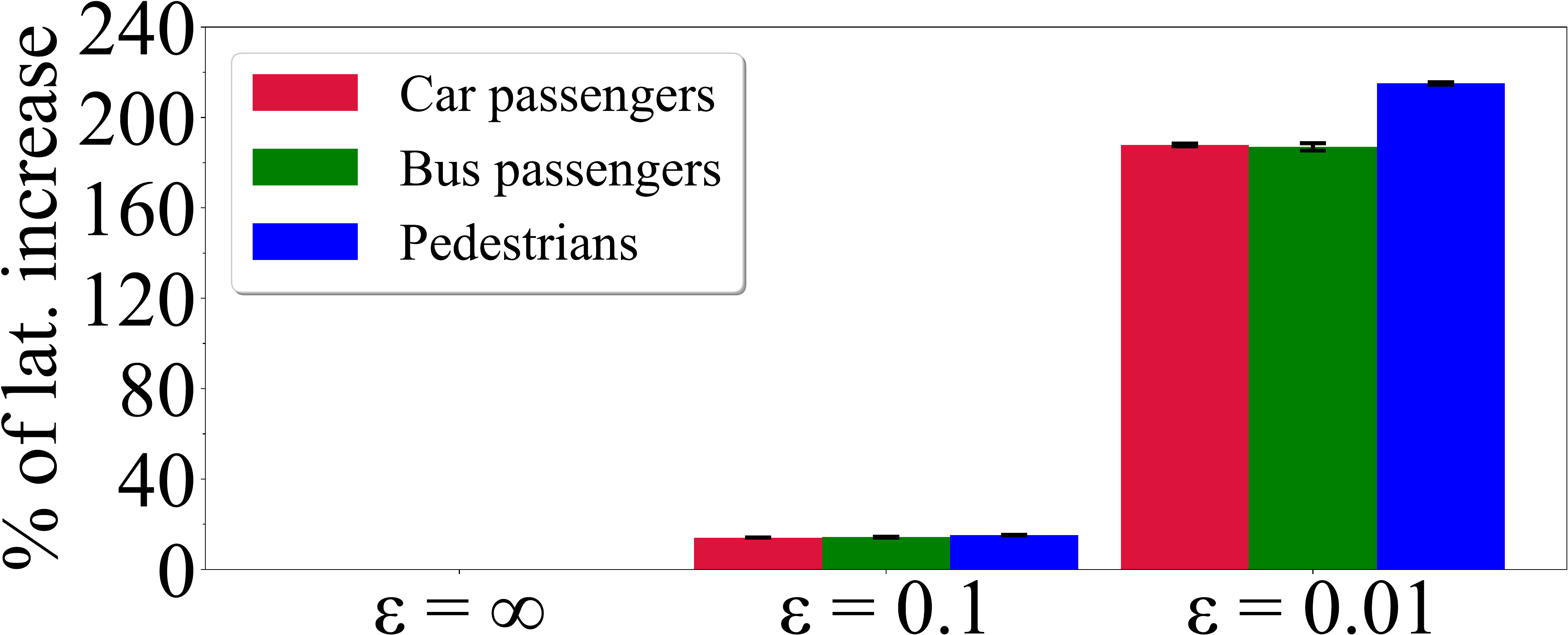}
 	\caption{Impact of privacy and mobility on latency increase.}
 	\label{fig:latencyIncrease}
\end{center}
\vspace{-0.3cm}
\end{figure}

Figure~\ref{fig:latencyIncrease} shows the latency increase caused by the non-ideal MH selections for each privacy level and mobility type. As shown, car and bus passengers had similar results as they have similar mobility patterns. However, pedestrians again suffered more the impact of privacy, with an average latency increase near 220\% for {\it high} privacy level. As argued in the previous section, even if the user is able to offload his application, the price of privacy protection comes in the form of increased latency which, as Figure~\ref{fig:latencyIncrease} shows, impacts pedestrian's mobility patterns the most.

\subsection{Impact of application type}

\begin{figure}[tt!!]
\begin{center}
 	\includegraphics[width=0.455\textwidth]{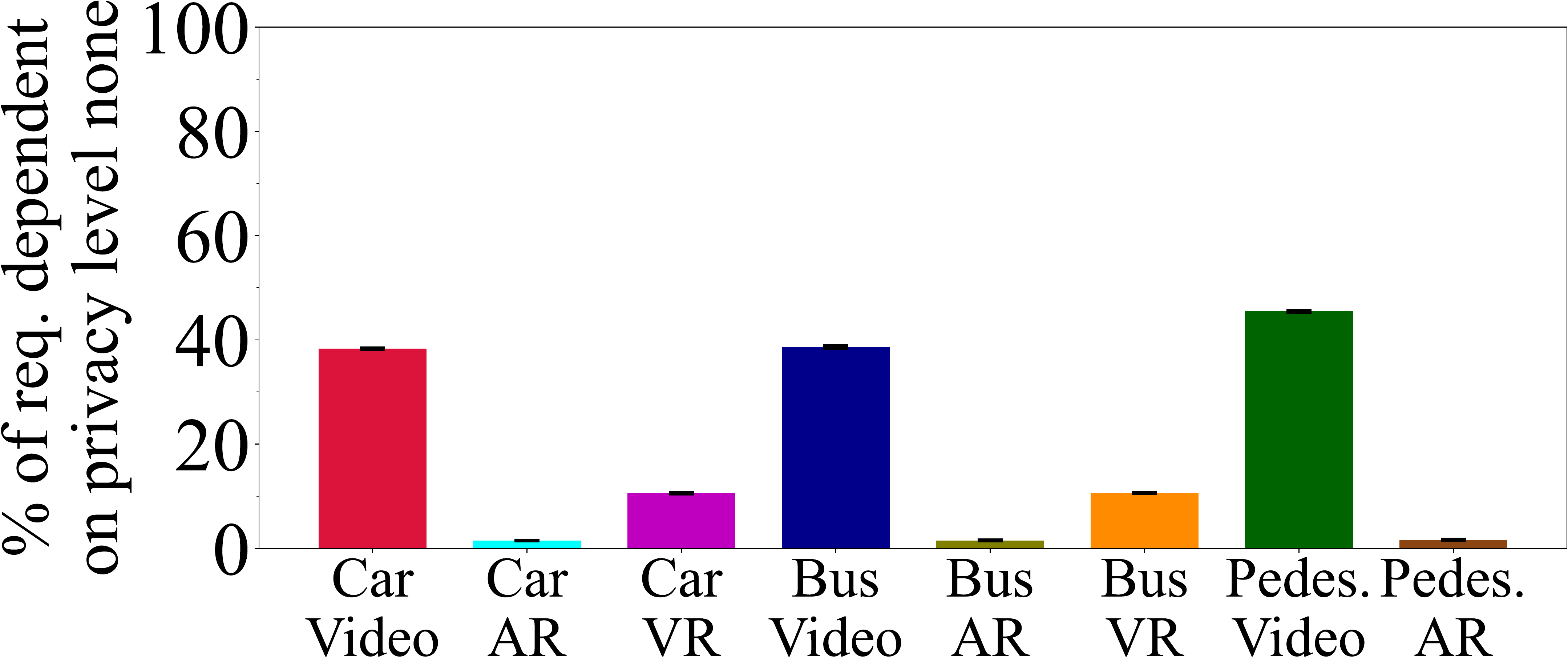}
 	\caption{Fraction of requests that succeeded when the privacy level was {\it none} but not on both remaining levels.}
 	\label{fig:privacyImpact}
\end{center}
\vspace{-0.3cm}
\end{figure}

Our last analysis focuses on the user requests that were successful (i.e., user could successfully offload his application) when the privacy level was set to {\it none} but not on both remaining levels, shedding light into how mobility and application of choice may impact users who could not achieve their desired privacy level. Figure~\ref{fig:privacyImpact} shows the fractions of those requests, computed over the total number of requests for each combination of mobility type and application type. As shown, for all mobility types, requests to the video streaming application, the most demanding in terms of latency, are the most affected by privacy protection. As already argued, the applications' latency requirement is more challenging to maintain than the bandwidth requirement, provided that MH capacity does not saturate. Moreover, once again we observe that pedestrians suffered more than car and bus passengers, emphasizing the importance of mobility patterns and how natural spatial constraints on mobility may yield different offloading performance.

Finally, we make source code and all data used in our experiments (including $\sim$405 million user requests on different privacy and mobility scenarios) publicly available for the sake of reproducibility and fostering future research\footnote{https://github.com/LABORA-INF-UFG/Offloading}.
    \section{Conclusions and future work}\label{secVI}

In this work, we analyzed the impacts of users' privacy and mobility when offloading to the edge. We carried out a large number of simulations based on multiple real-world scenarios, parameters, and applications, each with a specific user mobility model and privacy requirement scenario, in order to analyze the impacts that privacy and mobility can have. Additionally, we made publicly available code and data necessary for replicating our work. As future work, we plan to test different/adaptive privacy levels based on the UE's Quality of Service (QoS) and on the user's mobility pattern; analyze how different MH selection algorithms from the literature are affected by the privacy, and explore our findings in the design of novel MH selection algorithms that achieve competitive results while keeping pre-defined user privacy levels.

\section*{Acknowledgements}

The authors thank M\'{a}rio S. Alvim for his contributions and insights regarding privacy protection. This work was supported in part by \textit{Conselho Nacional de Desenvolvimento Cient\'{i}fico e Tecnol\'{o}gico} (CNPq), by \textit{Funda{\c c}\~ao de Amparo {\`a} Pesquisa do Estado de Minas Gerais} (FAPEMIG), by \textit{Coordena{\c c}\~ao de Aperfei{\c c}oamento de Pessoal de N\'{i}vel Superior} (CAPES) and in by the CAPES-STIC-AMSUD 22-STIC-07 LINT project.
    
    \bibliographystyle{ieeetr}
    \typeout{}
    \bibliography{utils/references.bib}

\begin{thebibliography}{10}

\bibitem{addad2020fast}
R.~A. Addad, D.~L.~C. Dutra, M.~Bagaa, T.~Taleb, and H.~Flinck, ``{Fast service
  migration in 5G trends and scenarios},'' {\em IEEE Network}, vol.~34, no.~2,
  pp.~92--98, 2020.

\bibitem{abbas2017mobile}
N.~Abbas, Y.~Zhang, A.~Taherkordi, and T.~Skeie, ``{Mobile edge computing: A
  survey},'' {\em IEEE Internet of Things Journal}, vol.~5, no.~1,
  pp.~450--465, 2017.

\bibitem{zhan2020mobility}
W.~Zhan, C.~Luo, G.~Min, C.~Wang, Q.~Zhu, and H.~Duan, ``{Mobility-aware
  multi-user offloading optimization for mobile edge computing},'' {\em IEEE
  Transactions on Vehicular Technology}, vol.~69, no.~3, pp.~3341--3356, 2020.

\bibitem{yang2019efficient}
C.~Yang, Y.~Liu, X.~Chen, W.~Zhong, and S.~Xie, ``{Efficient mobility-aware
  task offloading for vehicular edge computing networks},'' {\em IEEE Access},
  vol.~7, pp.~26652--26664, 2019.

\bibitem{he2017privacy}
X.~He, J.~Liu, R.~Jin, and H.~Dai, ``{Privacy-aware offloading in mobile-edge
  computing},'' in {\em GLOBECOM 2017-2017 IEEE Global Communications
  Conference}, pp.~1--6, IEEE, 2017.

\bibitem{sun2020reducing}
W.~Sun, H.~Zhang, R.~Wang, and Y.~Zhang, ``{Reducing offloading latency for
  digital twin edge networks in 6G},'' {\em IEEE Transactions on Vehicular
  Technology}, vol.~69, no.~10, pp.~12240--12251, 2020.

\bibitem{wei2021uav}
D.~Wei, N.~Xi, J.~Ma, and L.~He, ``{UAV-assisted privacy-preserving online
  computation offloading for internet of things},'' {\em Remote Sensing},
  vol.~13, no.~23, p.~4853, 2021.

\bibitem{peng2022mobility}
K.~Peng, P.~Liu, M.~Bilal, X.~Xu, and E.~Prezioso, ``{Mobility and
  Privacy-Aware Offloading of AR Applications for Healthcare Cyber-Physical
  Systems in Edge Computing},'' {\em IEEE Transactions on Network Science and
  Engineering}, 2022.

\bibitem{pang2022towards}
X.~Pang, Z.~Wang, J.~Li, R.~Zhou, J.~Ren, and Z.~Li, ``{Towards Online
  Privacy-preserving Computation Offloading in Mobile Edge Computing},'' in
  {\em IEEE INFOCOM 2022-IEEE Conference on Computer Communications},
  pp.~1179--1188, IEEE, 2022.

\bibitem{stallings2017data}
W.~Stallings, ``{Data and computer communications},'' 2017.

\bibitem{kim2005proportional}
H.~Kim and Y.~Han, ``{A proportional fair scheduling for multicarrier
  transmission systems},'' {\em IEEE Communications letters}, vol.~9, no.~3,
  pp.~210--212, 2005.

\bibitem{andres2013geo}
M.~E. Andr{\'e}s, N.~E. Bordenabe, K.~Chatzikokolakis, and C.~Palamidessi,
  ``{Geo-indistinguishability: Differential privacy for location-based
  systems},'' in {\em Proceedings of the 2013 ACM SIGSAC conference on Computer
  \& communications security}, pp.~901--914, 2013.

\bibitem{paverd2014modelling}
A.~Paverd, A.~Martin, and I.~Brown, ``{Modelling and automatically analysing
  privacy properties for honest-but-curious adversaries},'' {\em Tech. Rep},
  2014.

\bibitem{zhou2015spatial}
S.~Zhou, D.~Lee, B.~Leng, X.~Zhou, H.~Zhang, and Z.~Niu, ``{On the spatial
  distribution of base stations and its relation to the traffic density in
  cellular networks},'' {\em IEEE Access}, vol.~3, pp.~998--1010, 2015.

\bibitem{3gpp2016study}
3GPP, ``{Study on scenarios and requirements for next generation access
  technologies},'' {\em Technical Specification Group Radio Access Network,
  Technical Report 38.913}, 2016.

\bibitem{spinelli2020toward}
F.~Spinelli and V.~Mancuso, ``{Toward enabled industrial verticals in 5G: A
  survey on MEC-based approaches to provisioning and flexibility},'' {\em IEEE
  Communications Surveys \& Tutorials}, vol.~23, no.~1, pp.~596--630, 2020.

\bibitem{qiao2019effective}
Y.~Qiao, Z.~Liu, H.~Lv, M.~Li, Z.~Huang, Z.~Li, and W.~Liu, ``{An effective
  data privacy protection algorithm based on differential privacy in edge
  computing},'' {\em IEEE Access}, vol.~7, pp.~136203--136213, 2019.

\bibitem{lai2017furion}
Z.~Lai, Y.~C. Hu, Y.~Cui, L.~Sun, and N.~Dai, ``{Furion: Engineering
  high-quality immersive virtual reality on today's mobile devices},'' in {\em
  Proceedings of the 23rd Annual International Conference on Mobile Computing
  and Networking}, pp.~409--421, 2017.

\end{thebibliography}

\end{document}